%% file: IMC_SSW_2008_L3S_CR.tex
\documentclass[runningheads]{llncs}

\usepackage{amssymb,amsmath,bm,subfig,enumerate}
\usepackage{graphicx}

\usepackage{url}
\urldef{\mailsa}\path|{stewart,diaz,nejdl}@L3S.de|
\newcommand{\keywords}[1]{\par\addvspace\baselineskip
\noindent\keywordname\enspace\ignorespaces#1}

\begin{document}

\mainmatter  

\title{Mining User Profiles to Support Structure and Explanation in
  Open Social Networking}

\titlerunning{Mining User Profiles in OSN}

%
%
\author{Avar\'{e} Stewart \and Ernesto Diaz-Aviles \and Wolfgang Nejdl}
\authorrunning{Avar\'{e} Stewart \and Ernesto Diaz-Aviles \and Wolfgang Nejdl}

\institute{L3S Research Center / Leibniz Universit\"{a}t Hannover\\
  Appelstr. 9a, 30167 Hannover, Germany\\
  \mailsa\\
  \url{http://www.L3S.de/}}
%
%
\toctitle{Mining User Profiles to Support Structure and Explanation in Open Social Networking}
\tocauthor{Avar\'{e} Stewart, Ernesto Diaz-Aviles, Wolfgang Nejdl}
\maketitle

\begin{abstract}
  The proliferation of media sharing and social networking websites
  has brought with it vast collections of site-specific user generated
  content. The result is a Social Networking Divide in which the
  concepts and structure common across different sites are hidden. 
  
  The knowledge and structures from one social site are not adequately 
  exploited to provide new information and resources to the same or 
  different users in comparable social sites.  For music bloggers, this latent structure, forces bloggers to select sub-optimal blogrolls.  However, by integrating the social activities of music bloggers and listeners, we are able to overcome this limitation: improving the quality of the blogroll neighborhoods, in terms of similarity, by 85 percent when using tracks and by 120 percent when integrating tags from another site.   
  

\end{abstract}


\input{introduction}
\input{relatedwork}
\input{approach}

\input{experiment}
\input{conclusion}
\subsubsection*{Acknowledgements.}
This work was funded in part by the European Project PHAROS (IST Contract No. 045035), and by the Programme AlBan, the European Union Programme of High Level Scholarships for Latin America, scholarship no. (E07D400591SV).

\label{references}

\bibliographystyle{splncs}
\bibliography{ICM_SSW}

\end{document}

%% file: introduction.tex
%
%
\section{Introduction}
\label{sec:intro}
The increasingly growing collections of user generated content spread over heterogeneous social networking and media sharing platforms, each supporting specific media types. The content typically has a latent structure and latent, interrelated topics: this has resulted in a Social Networking Divide. 

Recent advances toward a more \emph{Open Social Networking}
(OSN) paradigm are focused on (\textit{de facto}) standards, and only address part of the problem. Specifically, current OSN efforts attempt to handle issues related to the portability of data, common APIs (e.g., Google OpenSocial\footnote{\url{http://code.google.com/opensocial}}), and social graphs, e.g., FOAF\footnote{\url{http://www.foaf-project.org/}}, XHTML Friends Network\footnote{\url{http://gmpg.org/xfn/}}.  

We posit that open social networking is more than an agreed upon ``language'' for describing relationships and sharing data across systems. In addition, it is the exploitation of social activities in one site, to support the discovery of new interrelationships within a community. This is crucial, given that it is becoming increasingly difficult for seekers to cope with the cognitive challenges of efficiently finding and effectively analyzing relevant information, when inundated with its volume, variety and evolution.

\subsection{Scenario}
\label{sec:scenario}
To motivate the aforementioned ideas, consider the following scenario in which there are two social network sites. In one site, a \emph{Blogger.com}\footnote{\url{http://www.blogger.com}}, music community, the main activities are writing text about artists, tracks, albums or music videos. Bloggers create explicit links to other participants to express their preferred blogs (via a $blogroll$). In the other site: \emph{Last.fm}\footnote{\url{http://www.last.fm}}, the users listen to music tracks, tag these tracks and build friendship relationships.

Symbiotically, the social activities in one site can have an impact on the other site. Blogger.com bloggers do not tag the entities about which they write. However, the tagging activity can better help bloggers see the structure in their community and find new information. Conversely, Last.fm users do not provide prose for the tracks they
listen to, but such prose can be a valuable source of metadata for audio tracks.

\begin{figure}[tbh]
  \centering
  \includegraphics[width=.7\textwidth]{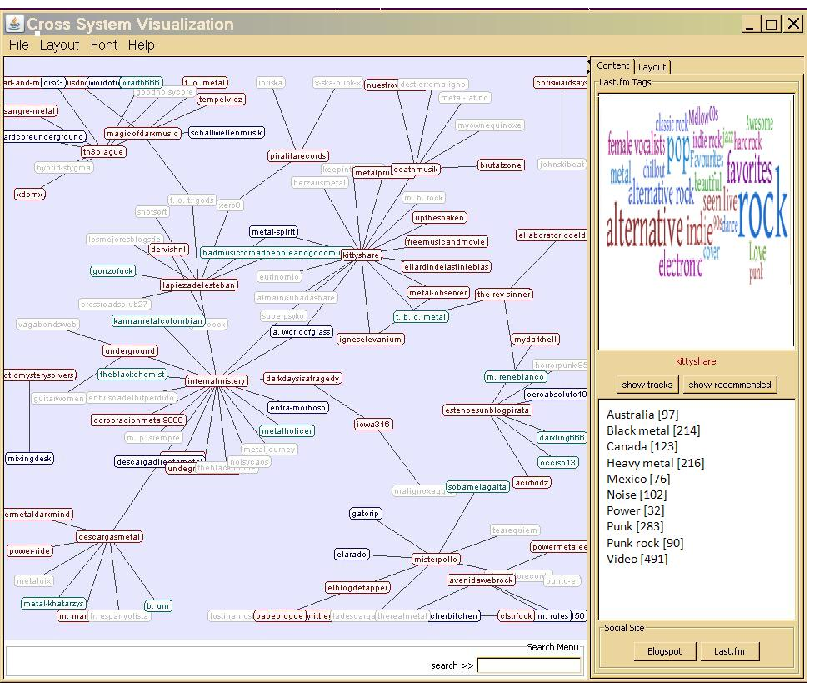}
  \caption{Cross System Visualization Blogger.com and Last.fm}
  \label{fig:crossVis}
\end{figure}

In Figure~\ref{fig:crossVis}, a navigation tool is depicted, in which the Blogger.com site has been enriched with information from Last.fm. The graph represents the similarity between (potentially unknown) bloggers  on the set of tracks they have written about in their blogs. By selecting a node in the graph, the list if tracks that blogger has mentioned in their blogsite is presented, along with the overall popularity of these tracks. Also depicted in the figure are the tags from Last.fm, which can
be used to filter the nodes and edges in the graph. The tool supports navigation and visualization of latent concepts and relations within, and across the sites.
One of the challenges in realizing such a scenario is that the concepts
and structure within ---and across domains--- are latent. Specifically
within blogs, the topics to which the blog site is devoted, are very
often not made explicit. Furthermore, the readership relationship is
typically unobserved; and the blogroll relationship, though observed,
is often unexplained. Finally, standards may support, but do not
address, how the social practices in one domain may be exploited to
support bloggers in another similar social site; particularly when
resources of different types are being mapped.

The contributions of this work are: 1) extension of the commonly held view to open social networking:  to infer new relationships and resources, and provide support with navigation and visualization across comparable social networks; 2) integration music bloggers with music listeners; bridging the gap between different types of social networking systems; 3) examination of the blogroll relationship in the music domain based an open social networking approach.
 
In Section~\ref{sec:relatedwork} we discuss related research in cross domain discovery. In Section~\ref{sec:approach} we present the conceptual approach to open social networking in the music domain. In Section~\ref{sec:experiments} experimental results are presented and in Section~\ref{sec:conclusion} we conclude and discuss future work.
%

%% file: relatedwork.tex
%
%
\section{Related Work}
\label{sec:relatedwork}
In machine learning, cross domain discovery~\cite{DBLP:conf/aaai/SwarupR06} or domain
adaptation~\cite{journals/jair/DaumeM06} is a body of work in which multiple
information sources from comparable, but different domains are
combined. Work done in this area has focused on classification tasks, in which labeled data exists in abundance in one domain, but a statistical model that performs well on a related domain is desired. Since hand-labeling in the new domain is costly, one often wishes to
leverage the original out-of-domain data when building a model of the
new, in-domain data~\cite{DaiXueYan+07}. 

Also in the area of machine learning, the method of \emph{View
Completion}~\cite{SBS08}, has used collaborative tags to heuristically
complete missing or inadequate feature sets (or views). The basic premise underlying view completion, is that for many tasks, combining multiple information sources yields significantly better
results than using just a single one alone. Views are used in this case, since that blogs are not typically available on collaborative tagging websites, and as such the tags provided by
bloggers suffer from the vocabulary problem and cannot be adequately used as a shared index.

Another related area is \emph{Cross-System
  Personalization}~\cite{MehHof07Man} which enables personalization
information across different systems to be shared. In digital
libraries,  cross system personalization is used to overcome the problem
that information needed to support a personalized user experience is
not shared among different libraries.  In other work, the focus has been on adequate representations
of~\cite{NieSteMehHem+2004,WanHhaZha07}, and dependencies
between~\cite{MehHof07Fac} the user's profile, to support a
unified representation in the different systems. These approaches are ego-centric in that they assume the same user 
to exist across different systems; and that the user is interested in
an aggregated view of their profile or social networking information.
This is not the case in an open social networking environment, where
users are assumed to be similar (in some way), but have distinct
digital identities.  Our view is a socio-centric one and focuses on common patterns in the community as a whole.

Work exists, in the area of association mining~\cite{Oldenburg08,JasHotChr+08}. In ~\cite{Oldenburg08} the goal is to provide a seamless navigation between tag spaces.
The work presented in~\cite{JasHotChr+08} merges the
areas of formal concept analysis and association rule mining to
discover shared conceptualizations that are hidden in folksonomies. They present a formalism for folksonomies that includes a set $U$ of users, a set $T$ of tags, and a set $R$ of resources, represented by the ternary relation, $Y$. In our work, it is the set, $T$, of tags in the ternary relation that
we propagate from one site to enrich another. Furthermore, we propose, that
``citizen-defined'' structuring (i.e. blogroll, friends, or comment
networks) allow other types of ternary relations to be inferred, that
are not restricted to the \textit{user-tag-resource} triple.

%% file: approach.tex
%
%
\section{Open Social Networking: the Music Domain}
\label{sec:approach}
Open social networking, has two-fold goal: 1)improve the structure of information within a single site and 2)exploit the social activities in a different sites to enhance the activities in comparable ones. Toward this end, two aspects are considered: a representation for the activities within each social site; and mapping  parts of the data and structures from one social site to another, to augment the social activities therein.

\subsection{Terminology}
We adopt the definition of a folksonomy as described by Hotho~et~al.~\cite{JasHotChr+08,Hotho-Information_2006}, as a four-tuple\footnote{In the original definition \cite{JasHotChr+08,Hotho-Information_2006}, it is additionally introduced a subtag/supertag relation, which
we omit for the purpose of this paper.}, 
$\mathbb{F} := (U,T,R,Y)$ , where:
\begin{itemize}
	\item[$\bullet$] $U$,$T$ and $R$ are finite sets, whose elements are called users, tags and resources, respectively, and
	\item[$\bullet$] $Y$ a ternary relation between them, i.e. $Y \subseteq U \times T \times R$, whose elements are called tag assignments.
\end{itemize}

For the music domain, the set of resources are considered to be artists, tracks and albums. Additionally, we distinguish the different roles a site may have when describing the mappings between them. A target site, or \emph{in-domain site}, is the one onto which data from another social site is mapped. The \emph{out-of-domain site} is the social site from which data is extracted to augment the comparable, target site. The roles of in- and out-of-domain may be interchanged depending upon integration goals, and there may be multiple out-of-domain sites. For the purpose of this work, we consider $Blogger.com$ and $Last.fm$ to be the \emph{in-domain site} and \emph{out-of-domain site}, respectively. 

\subsection{Cross-Site Enrichment}
We represent the Blogger.com music community conceptually as tuples, $B := \{(u_b,r_b)\;|\;(u_b,r_b) \in U_{Blogger.com} \times R_{Blogger.com}\}$, since Blogger.com bloggers do not tag the entities about which they write. On the other hand, the Last.fm social site is ripe with tag data of the form $L := \{(u_l,t_l,r_l)\;|\;(u_l,t_l,r_l) \in U_{Last.fm} \times T_{Last.fm} \times R_{Last.fm}\}$, representing the tag a given user has applied to a track within Last.fm. Then the mapping of tags onto Blogger.com is computed as follows:

\begin{equation}
	Y := \{(u,t,r)\;|\;(u,t,r) \in \pi_{u_b,t_l,r_b}(\sigma_{r_b=r_l}(B \times L)) \}
\end{equation}

were $\sigma$ and $\pi$ are the relational algebra operators for selection and projection, respectively. 

First, from the cartesian product $B \times L$, the tuples with equal resources in both sites are selected, and then the projection is taken over the Blogger.com users, the Last.fm tags, and the common resource elements. In general, the user sets are considered to be disjoint, i.e., $U_{Blogger.com} \cap U_{Last.fm} = \emptyset$.

\subsection{Site-Specific Enrichment}
The (hidden) relationships between blogs and/or bloggers can be exploited to infer relationships between the entities within the blog site. However, even within a single blog community, the relationships between resources, may not be well understood. For this reason, we undertake an exploratory analysis of the blogroll relationship.

%% file: experiment.tex
%
%
\section{Experiments}
\label{sec:experiments}
The experimental goals are to first examine the explicit blogroll structure; laying the foundation for further analysis of an ideal or ``optimal'' resource-specific blogrolls. Resource specific blogrolls are those in which the nature of the blogroll is assumed to be explained in terms similarity in tastes for a given type of resource, i.e., track, or artist. Then, to investigate the extent to which these optimal resource-specific blogrolls: overlap with the explicit blogrolls;  and with each other. In the remainder of this discussion "`optimal"' resource-specific blogrolls is referred to as optimal blogrolls.

\subsection{Data Set}
For Cross System Music Blog Mining, we used two data sets: one data set consisted of personal music blogs from $Blogger.com$, one of the most popular blogsites, whereas the second data set consisted of tagged tracks from $Last.fm$, a radio and music community website and one of the largest social music platforms. The details of each data set are presented in this section.\\
\textit{Blogger.com Community:} The blogroll relationship induces a network representing a preferential reading of others people's blogs. The network data was collected by experimentally selecting seed bloggers using several music directories\footnote{\url{http://www.musicblogscatalog.com/}\\\url{http://yocheckthisjam.com/music-blog-directory/}\\\url{http://www.blogged.com/directory/entertainment/music/rock}\\\url{http://www.blogcatalog.com/directory/music/rock}} and limiting the bloggers selected to the genre of pop and rock music in the Blogger.com domain. The blogroll for each seed was traversed, fanning out in a breath-first order, yielding a total number of bloggers equal to $|U_{Blogger.com}|=976$.

Summary statistics for the overall structure and topological statistics for the largest five weak components are given in Table~\ref{tab:Blogger.comCommunityOverallStatistics}, from there it can be seen that the components exhibit varying structural properties and that the structural view provided by the blogroll is a disjointed one.

In addition to the community data, profiles were built by parsing the tracks in the user's blog and relying upon a dictionary of tracks gathered from MusicBrainz.org\footnote{\url{http://www.musicbrainz.org}}. A total of 2196 unique tracks were collected; and for these tracks, a total 147801 Last.fm tags were obtained, which allowed us to construct the triples.
\begin{table}
\begin{center}
\caption{Blogger.com Community Statistics and Components}
\label{tab:Blogger.comCommunityOverallStatistics}
\begin{tabular}{|l|l|l|l|l|l|l|}
\hline\hline
	  &   &    & Avg.   & Max.  & Min.  &    \\
	No. & No. & Weak  &   Component  &   Component &   Component& Reciprocal  \\
	Nodes & Edges &  Components &  Size & Size & Size & Edges \\ [0.5ex]
\hline
	976 & 2011 & 72 & 13.55 & 662 & 2 & 581 \\ [0.5ex]
\hline
\hline\hline
	      & & & Avg. & Shortest  & Longest   &  \\
      Comp.   & No. & No. & Cluster & Average & Average & Reciprocal \\
      Id & Nodes &  Edges & Coef. & Distance  & Distance & Edges\\[0.5ex]
\hline
      1 & 662 &  1755 & 0.492 & 2.797 &8.134 & 568 \\
      2 & 53  &  62 & 0.574 & 2.113 & 5.811 & 13 \\
      3 & 28  &  27 & 0.964 & 0.964 & 1.892 & 0 \\
      4 & 19  &  20 & 0.526 & 2.0 & 4.684 & 0 \\
      5 & 16  &  15 & 0.751 &1.687 & 3.187 & 0 \\[0.5ex]
\hline\hline
\end{tabular}
\end{center}
\end{table}
\subsection{Experimental Results}
\subsubsection{Blogroll Quality Based on Tracks and Tags} 
We investigate the extent to which the explicit blogroll relationship, within the in-domain site, can be described by the similarity between track and tag profiles. For each user $u \in U$, i.e. blogger, a track-based vector profile (\textit{track profile}) $\textbf{u}$ is constructed such that $\textbf{u} := \{0,1\}^{|R|}$, with the ith dimension $\textbf{u}_i$ set to 1 if the track $r_i \in R_u$, appears in the user's blog, and 0 otherwise.\\
Alternatively, after including the tag information from the out-of-domain site, we constructed a profile based on tag annotations, which corresponds to the \textit{tag   profile} for the user, i.e., $\textbf{u} := \{0,1\}^{|T|}$, with the ith dimension $\textbf{u}_i$ set to 1 if the tag $t_i \in T_u$, and 0 otherwise\footnote{The 20000 most popular tags were used to build the profiles, i.e. $|T|=20000$}.\\
To evaluate the quality of the explicit blogroll $B_u$ of user $u$, we computed an average similarity score between a user and all the persons in his blogroll. We perform the similarity computation using the cosine-based measure:

\begin{equation}
	sim(\textbf{u},\textbf{v}) := cos(\textbf{u},\textbf{v}) := \frac{\langle \textbf{u},\textbf{v} \rangle}{||\textbf{u}|| \cdot ||\textbf{v}||} 
\end{equation}

where $\textbf{u}$ and $\textbf{v}$ denote either the track or tag  profiles of users $u$ and $v$, respectively.

For the optimal blogroll $B^*$ computation, we constructed a user-track (resp., user-tag) profile matrix and proceeded as follows:

\begin{enumerate}[(i)]
	\item Compute the user similarity matrix $S_{|U|\times|U|} := (sim(\textbf{u}, \textbf{v}))$
	\item Keep the highest $k$ entries in each column of $S$.
	\item Set the optimal blogroll based on track profile (resp., tag profiles) for user $u$, i.e., $B^*_{u\;R}$ (resp., $B^*_{u\;T}$), to be the users in the non-zero columns of the corresponding $u$'s row.
\end{enumerate}
In our experiments we set the value of parameter $k = 10$. Table~\ref{tab:experimentalResults} summarizes the results.
\begin{table}[t]
  \begin{minipage}[t]{1\linewidth}
    \centering
    \caption{Average similarity and overlap of the blogroll $B$ and optimal blogroll $B^*$}
    \label{tab:experimentalResults}
    \begin{tabular}{|l|l|l|l|p{3.5cm}|}
\hline\hline
      Profile &  $AvgSim(B)$ & $AvgSim(B^*)$ & Improvement ($\%$) & $Avg(|B \cap B^*|)$ \\[0.5ex]
      \hline
	Track-Based & $0.295$ & $0.547$ &  $85\%$ &  $1.48 \approx 1$ blogger, with probability $= 0.085$ \\
	Tag-Based   & $0.293$ & $0.645$ & $120\%$  & $1.37 \approx 1$ blogger, with probability $= 0.081$\\[0.5ex]
\hline\hline	
    \end{tabular}
  \end{minipage}
\end{table}
\subsubsection{$B^*_R$ and $B^*_T$ Overlap}
\begin{figure}[t]
    \centering
    \includegraphics[width=0.9\linewidth]{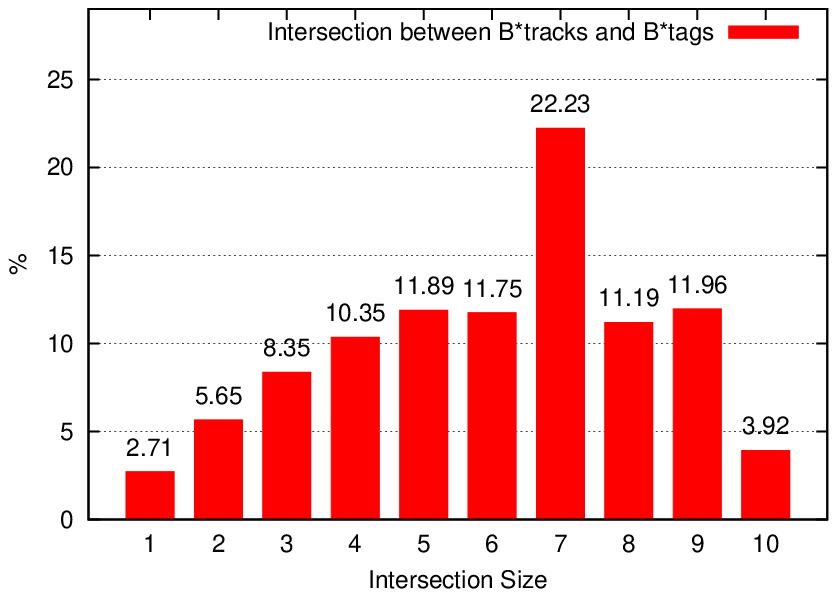}
    \caption{Distribution of intersection size between the optimals blogrolls $|B^*_R \cap B^*_T|$}
  \label{fig:intersectionSize}
\end{figure}

measures the extent to which optimal blogrolls, computed with track and tag profiles, agree on his members. We found that $77.66\%$ of the time they agree on at least one member, and the average of the overlap in these cases is $Avg(|B^*_{R} \cap B^*_{T}|)=4.64 \approx 5$ bloggers out of $10$ (the fixed size of the optimal blogrolls). The distribution of the intersection size for the optimal blogrolls is presented in Fig.~\ref{fig:intersectionSize}.
\begin{figure}[t]
  \centering
  \subfloat[]{\label{fig:cumuFreqTrack}\includegraphics[width=0.5\textwidth]{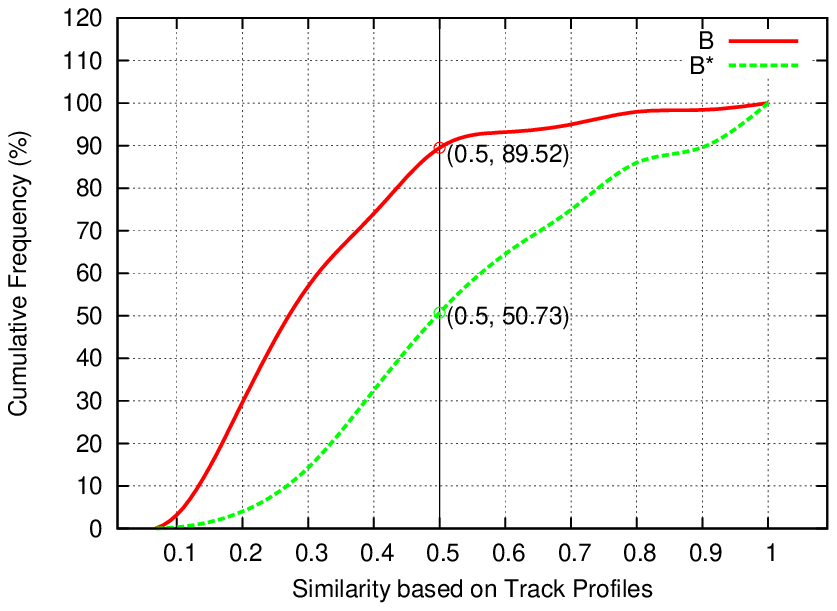}}                
  \subfloat[]{\label{fig:cumuFreqTag}\includegraphics[width=0.5\textwidth]{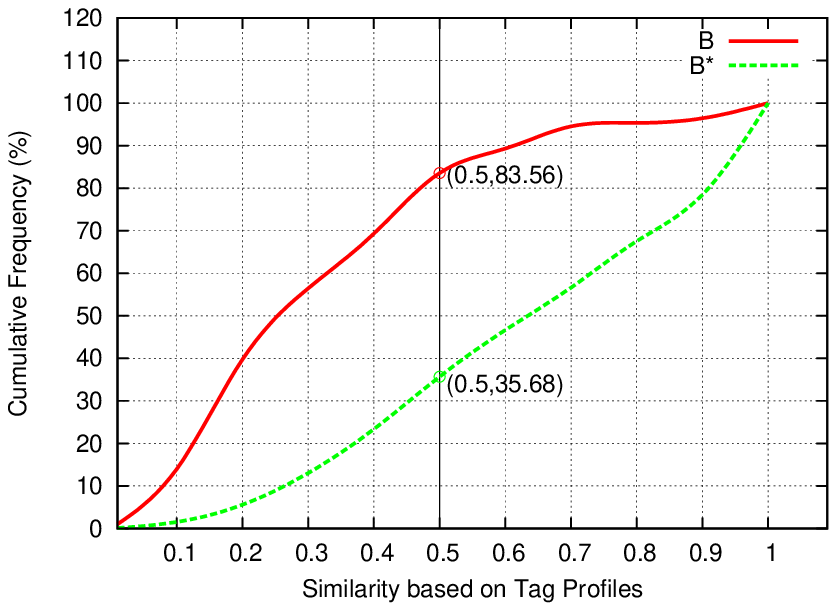}}
  \caption{Cumulative Frequency for the explicit blogroll $B$ and the optimal $B^*$ computed using (a) Track Profiles and (b) Tag Profiles.}
  \label{fig:cumuFreq}
\end{figure}
\begin{figure}[t]
  \centering
  \subfloat[]{\label{fig:freqTrack}\includegraphics[width=0.5\textwidth]{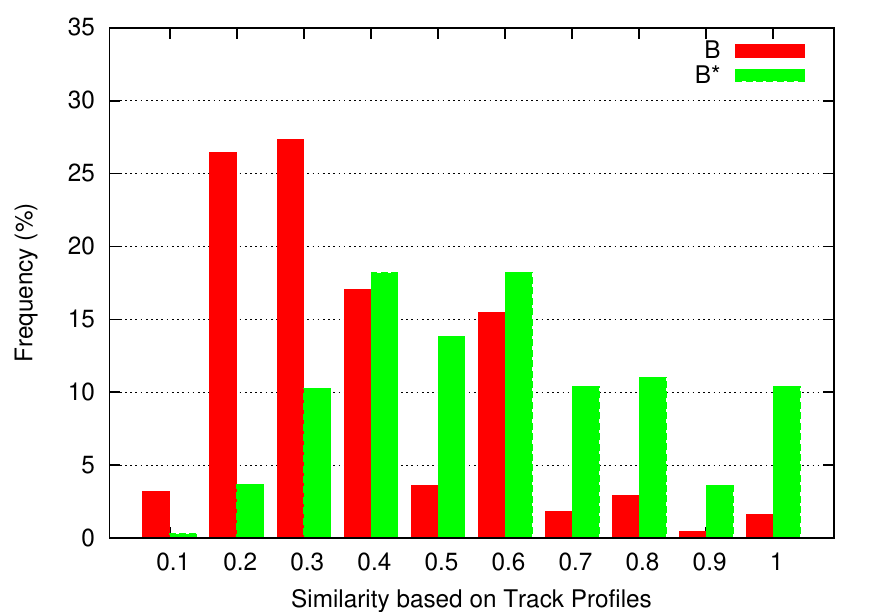}}                
  \subfloat[]{\label{fig:freqTag}\includegraphics[width=0.5\textwidth]{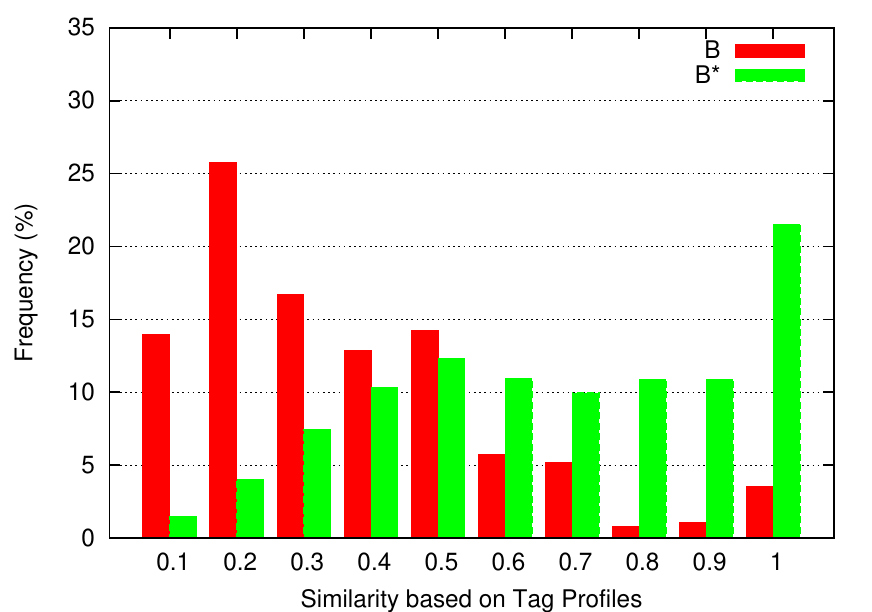}}
  \caption{Frequency Distribution for the explicit blogroll $B$ and the optimal $B^*$ computed using (a) Track Profiles and (b) Tag Profiles.}
  \label{fig:freq}
\end{figure}
\subsection{Discussion}
From Table~\ref{tab:experimentalResults}, it can be observed that the improvement, in terms of similarity, when computing $B^*$ based on tracks is $85\%$, and $120\%$ when tag profiles are used. The table also shows that the overlap between the explicit and optimal blogrolls, computed either with track or tag profiles, occurs only $9\%$ of the time, corresponding to an average of a single blogger in those cases.

Furthermore, the optimal blogroll similarity distributions are better than the one produced by the explicit relationships, as shown in Fig.~\ref{fig:freq}, which corresponds to the absolute frequencies of the explicit and optimal blogrols, for different values of similarity. The frequencies of optimal blogrolls, for similarity values over $0.5$, are higher than the ones for explicit blogrolls.

The cumulative frequency of blogrolls over discrete similarity bins is presented in Fig.~\ref{fig:cumuFreq}, which shows that both the track-based ($49.27\%$) and tag-based ($64.32\%$) optimal blogrolls exhibit good similarity quality, i.e., over $0.50$, in contrast to the respective explicit blogrolls, where just less than $11\%$ of them in the case of track-based profiles (resp., $16.44\%$ for tag-based) fall in bins corresponding to similarity values over $0.5$.\\
Tag based computations perform better than using tracks, i.e., builds optimal blogrolls with higher values of similarity, this can be explained by the fact that tags capture some structure of the domain, e.g., genre, improving the overlap between user profiles when computing the similarity measure.
%
%

%% file: conclusion.tex
%
%
\section{Conclusions}
\label{sec:conclusion}
In this paper, we explored to what extent the knowledge and structures from one social site (out-of-domain) can be adequately exploited to provide new information and resources to the  users in a comparable social site (in-domain), in particular for a music blog community within the $Blogger.com$ social network. An examination of the explicit blogroll structure, which is assumed to express a preferential reading of others people's blogs, has revealed that bloggers tend to produce sub-optimal blogrolls when measuring the similarity between users based on track, as well as tags. The implication for this is that if users are interested in learning about tracks, tags or other bloggers, some assistance to guide them is needed. On the other hand, neither tracks nor tags comes close to fully ``explaining'' the nature of the blogroll.\\
However, by integrating the social activities of music bloggers and listeners, we were able to overcome this limitation. We have shown the improvement that Open Social Networking can have on the quality of blogrolls: Last.fm offers better optimal blogrolls, than the tracks alone, from Blogger.com, improving the quality of the blogroll neighborhoods, in terms of similarity, by $85\%$ when using tracks and by $120\%$ percent when integrating tags from the site. The higher value of similarities computed based on tags can be explained by the fact that tags capture some structure of the domain, e.g., genre, increasing the overlap probability and size on the user profiles. Though this do not necessarily mean that tags are better predictor of similarity than tracks, it strongly suggests that the kind of information captured by tags can be exploited effectively, complementing tasks or models where tracks are used alone.\\
Although our investigation has provided promising results, we believe that our contribution is an initial step in the study of Open Social Networking, future work is required to evaluate the usefulness of optimal blogrolls, e.g., in providing recommendations. Furthermore, we plan to investigate the extent to which the explicit community bonds and ``citizen-defined'' structuring (i.e. blogroll, friends, or comment networks) can be described by mining and inferring associations between the profiles of users across social sites, towards a more general model, that considers new dimensions beyond the ternary relation between users, tags and resources.